\documentclass[a4paper]{panl}
\usepackage{cite}
\usepackage{wrapfig}
\usepackage{graphicx}
\usepackage{amssymb}
\usepackage{amsfonts}
\usepackage{amsmath}
\usepackage{longtable}
\usepackage{rotating}
\usepackage{lscape}
\usepackage{epsfig}
\usepackage{multirow}

\originalTeX

\newcommand{\beq}{\begin{equation}}
\newcommand{\eeq}{\end{equation}}
\newcommand{\bea}{\begin{eqnarray}}
\newcommand{\eea}{\end{eqnarray}}

\newcommand{\eq}{\begin{equation}}
\newcommand{\en}{\end{equation}}
\newcommand{\eqa}{\begin{eqnarray}}
\newcommand{\ena}{\end{eqnarray}}
\newcommand{\half}{\frac{1}{2}}

\begin{document}

\title{ Abelian and monopole dominance in SU(3) gluodynamics and Gribov copy effects}
\maketitle
\authors{I.\,Kudrov$^{a,}$\footnote{E-mail: Ilya.Kudrov@ihep.ru},
V.\,Bornyakov$^{a,b,c,}$}
\setcounter{footnote}{0}
\from{$^{a}$\,NRC ``Kurchatov Institute'' - IHEP, Protvino, 142281 Russia}
\from{$^{b}$\,Pacific Quantum Center, Far Eastern Federal University, 690950 Vladivostok, Russia}
\from{$^{c}$\,KKTEP, NRC “Kurchatov Institute”,  Moscow, Russia}

\begin{abstract}
We continue our study of the Gribov copies effrcts in the Maximal Abelian gauge in lattice $SU(3)$ gluodynamics. Our computations were completed for four values of the lattice spacing with physical lattice size $L \approx 2$ fm. It is demonstrated that when one uses the effective simulated annealing algorithm to fix the gauge the obtained Gribov copies 
produce low  abelian string tension which is below  90\% of the physical value independent of the lattice spacing. These Gribov copies produce also low value (about 86\%) for the monopole string tension. It is further shown that in case of less effective relaxation algorithm it is possible to obtain Gribov copies which produce both Abelian and monopole string tension in good agreement with the physical one.
\\

\end{abstract}
\vspace*{6pt}


\label{sec:intro}
\section*{Introduction}

In this work, we carried out a study of the Gribov copiy effects in the Maximal Abelian (MA) gauge. This gauge is intensively used to study the dual superconductor scenario of confinement suggested in \cite{thooft,Mandelstam:1974pi}. The idea of this scenario is that condensation of  color-magnetic monopoles gives rise to squeezing of the color-electric flux directed from a static quark toward static anti-quark into a thin flux tube in the way analogous (but dual) to type II  superconductor. There are no classical monopole solutions in QCD or in gluodynamics and t'Hooft suggested to fix the Abelian gauge breaking $SU(N_c)$ to $U(1)^{N_c-1}$ to introduce Abelian monopoles as singularities of this gauge. Later on he introduced
MA gauge  \cite{thooft2} with the same purpose. 

The gauge fixing functional for MA gauge in case of $N_c=3$ considered here is the following
\beq
F =  \frac{1}{V} \int d^4 x  \sum_{\mu, a \neq 3,8} A^a_\mu(x) A^a_\mu(x),
\label{functional}
\eeq
where  $A^a_\mu(x)$ is a gauge field. This functional is invariant with respect to Abelian gauge transformation 
$g(x) \in U(1) \times U(1)$.  Respective gauge condition in the differential form satisfied by extrema of the functional (\ref{functional}) is nonlinear:
\begin{equation}
     f^a(A) = \sum\limits_{b \neq 3, 8}(\partial_{\mu} \delta^{ab} - g f^{ab3} A_\mu^3 - g f^{ab8} A_\mu^8) A_\mu^b =0\,,~~~~~~~~ a \neq 3, 8 
\end{equation}

MA gauge in lattice regularization was formulated in \cite{Kronfeld:1987vd,Brandstater:1991sn}. 
In this regularization the gauge fixing functional is of the form:
\begin{equation}
F_{lat} = 
1-\frac{1}{12\,V}
 \sum_{x,\mu}\left[ |U^{(11)}_\mu(x)|^2 +|U^{(22)}_\mu(x)|^2 
            +|U^{(33)}_\mu(x)|^2 \right] \,,
\label{functional2}
\end{equation}
where $U_\mu(x) \in SU(3)$ denotes lattice link gauge field.
Properties of MA gauge were intensively studied and results obtained supported the dual superconductor scenario of confinement, see Refs.~\cite{review1,review3,review4} for reviews. 

Functional (\ref{functional})  has numerous local minima corresponding to Gribov copies discovered by Gribov for the Coulomb gauge in Ref.~\cite{Gribov:1977wm}. Gribov's statement was generalized to other gauges in \cite{Singer:1978dk}.
In the framework of perturbation theory, this problem does not manifest itself and quantization can be successfully performed using the Faddeev-Popov method \cite{Faddeev:1967fc}. However, in the nonperturbative region, the Faddeev-Popov method does not work, since there are many gauge-equivalent configurations, called Gribov copies, satisfying a given gauge condition. 

Lattice regularization allows to study numerically nonperturbative properties of the nonabelian gauge theories, in particular the Gribov copy effects. In the MA gauge, strong Gribov copy effects were found, i.e., a strong dependence of gauge noninvariant observables on the choice of Gribov copies \cite{Bali:1996zs}. In  practice it is impossible to find global minima of the gauge functional numerically, but it is natural to assume that by generating a set of such minima and taking the minimal of them, we approach the global minimum. Such a practical approach to reduction of the Gribov copy effects was proposed in \cite{Bali:1996zs}, where the MA gauge was studied in lattice $SU(2)$   gluodynamics. This approach combined with effective gauge fixing algorithm was then used in studies of the MA gauge in both gluodynamics \cite{Bali:1996dm} and QCD \cite{DIK:2003alb}, as well as in studies of Landau gauge \cite{Bogolubsky:2005wf, Bornyakov:2008yx}, Coulomb gauge \cite{Burgio:2016nad} and center gauges \cite{DelDebbio:1998luz,Bornyakov:2000ig}.

MA gauge is used as a tool to locate the color-magnetic monopoles which are gauge invariant as was argued in \cite{Bonati:2010bb}. For this reason the proper Gribov copy might be different from the global minimum of the functional  (\ref{functional}). Similarly, center gauges are used to locate the gauge invariant center vorticies and it is a difficult task to
find proper Gribov copy in these gauges \cite{DelDebbio:1998luz,Golubich:2020sqd,Dehghan:2024rly}.
In our previous work \cite{Kudrov2024} we demonstrated that for $SU(3)$ gluodynamics it is possible to choose Gribov copies of MA gauge which allow
to obtain Abelian string tension very close to the physical value. Here we show more results confirming this conclusion and extend our study to the monopole dominance.

\section{Details of simulations and definitions}
Our simulations were done on lattices with lattice spacing $a$ varying between
0.093~fm and 0.059~fm with lattice size $L \approx 2$~fm, more details about our simulations can be found in  \cite{Kudrov2024}. For maximal lattice spacing ($a=0.093$ fm) we considered three lattice sizes to study finite volume effects.
To find the local minima (i.e. Gribov copies) of the functional (\ref{functional2}) we used two different numerical algorithms. The first one is relaxation supplemented by overrelaxation (RO). The second algorithm, simulated annealing (SA), is a much more efficient algorithm that is applied before the relaxation algorithm. This algorithm
produces local minima with lower values of the gauge fixing functional. 

After fixing the MA gauge we  performed the Abelian projection
which starts from a coset decomposition with respect to the subgroup $U(1) \times U(1)$
of the non-Abelian gauge field $U_\mu(x) \in SU(3)$   as defined in Ref.~\cite{Brandstater:1991sn}:
\begin{equation}
U_\mu(x) = U^{offd}_\mu(x) U^{Abel}_\mu(x) \,.
\end{equation} 
where $U^{offd}_\mu(x) \in SU(3)/U(1) \times U(1)$ is the off-diagonal component and $U^{Abel}_\mu(x) \in U(1) \times U(1)$
is the diagonal component. 
The Abelian field $U^{Abel}_\mu(x)$ is defined by relations  \cite{Brandstater:1991sn}:
\beq
\label{uabel} 
U^{Abel}_\mu(x) = \mbox{diag} \left(u^{(1)}_\mu(x),u^{(2)}_\mu(x),u^{(3)}_\mu(x) \right)\,,~~~~ 
u^{(a)}_{\mu}(x)=e^{i \theta^{(a)}_{\mu}(x)}\,,
\eeq 
\beq 
\label{tlink} 
\hspace{-2mm}
\theta^{(a)}_{\mu}(x) = \arg~(U_\mu^{aa}(x))-\frac{1}{3} \sum_{b=1}^3 \arg(U_\mu^{bb}(x))\,\big|_{\,{\rm mod}\ 2\pi}\,,~\theta^{(a)}_{\mu}(x) \in [-\frac{4}{3}\pi, \frac{4}{3}\pi]\,.
\eeq 

\section{Abelian dominance}

Abelian dominance was first observed in Ref.~\cite{suzuki1}. It was shown that in $SU(2)$ gluodynamics the Abelian string tension $\sigma_{ab}$ extracted from the Abelian Wilson loops after fixing MA gauge is approximately equal to the physical string tension $\sigma$. It was shown in Ref.~\cite{bm} that in that theory $\sigma_{ab}/ \sigma =1$ within error bars in the continuum limit. This ratio in $SU(3)$ gluodynamics was studied in Refs.~\cite{DIK:2003alb,Stack:2002ysv,Sakumichi:2014xpa} 
It was found in Refs.~\cite{DIK:2003alb,Stack:2002ysv}
that in this theory $\sigma_{ab}/ \sigma$ is substantially 
lower than 1 at least at $\beta=6.0$. Later on 
it was claimed in Ref.~\cite{Sakumichi:2014xpa} that the perfect Abelian dominance can be observed in $SU(3)$ gluodynamics on large enough lattices. 
In our recent work \cite{Kudrov2024} we demonstrated that to get $\sigma_{ab}/ \sigma$ close to $1$ one has to choose 
proper Gribov copies rather than to get rid of finite volume effects.  Below we present more evidence to support this conclusion.

Finally, in Fig.~\ref{poten_compar} we compare the static potentials $V(r)$ computed for original nonabelian gauge field $U_{\mu}(x)$ and for Abelian fields  $U^{abel}_{\mu}(x)$. The Abelian fields were obtained either for one gauge copy with RO algorithm (corresponding to the highest value of $F_{lat}$) or for $n=20$ gauge copies with SA algorithm (corresponding to the lowest value of $F_{lat}$). Results are shown for all four values of the lattice spacing. 
The potentials $V(r)$ and the distance $r$ are normalized by Sommer
parameter $r_0$ \cite{Sommer:1993ce}, the additive divergence in  $V(r)$ is removed by subtracting $V(r_0/2)$. All three potentials demonstrate weak dependence on the UV cutoff, i.e. on the lattice spacing.
The potentials were fitted to the Cornell potential and the ratio $\sigma_{ab}/ \sigma$ was obtained equal to 0.83(2) for SA algorithm (in agreement with Refs.~\cite{DIK:2003alb,Stack:2002ysv}) and to 0.96(3) for RO algorithm in agreement with  \cite{Sakumichi:2014xpa}.
\begin{figure}[t]
\vspace*{-1.5cm}
\begin{center}
\includegraphics[width=0.8\textwidth]{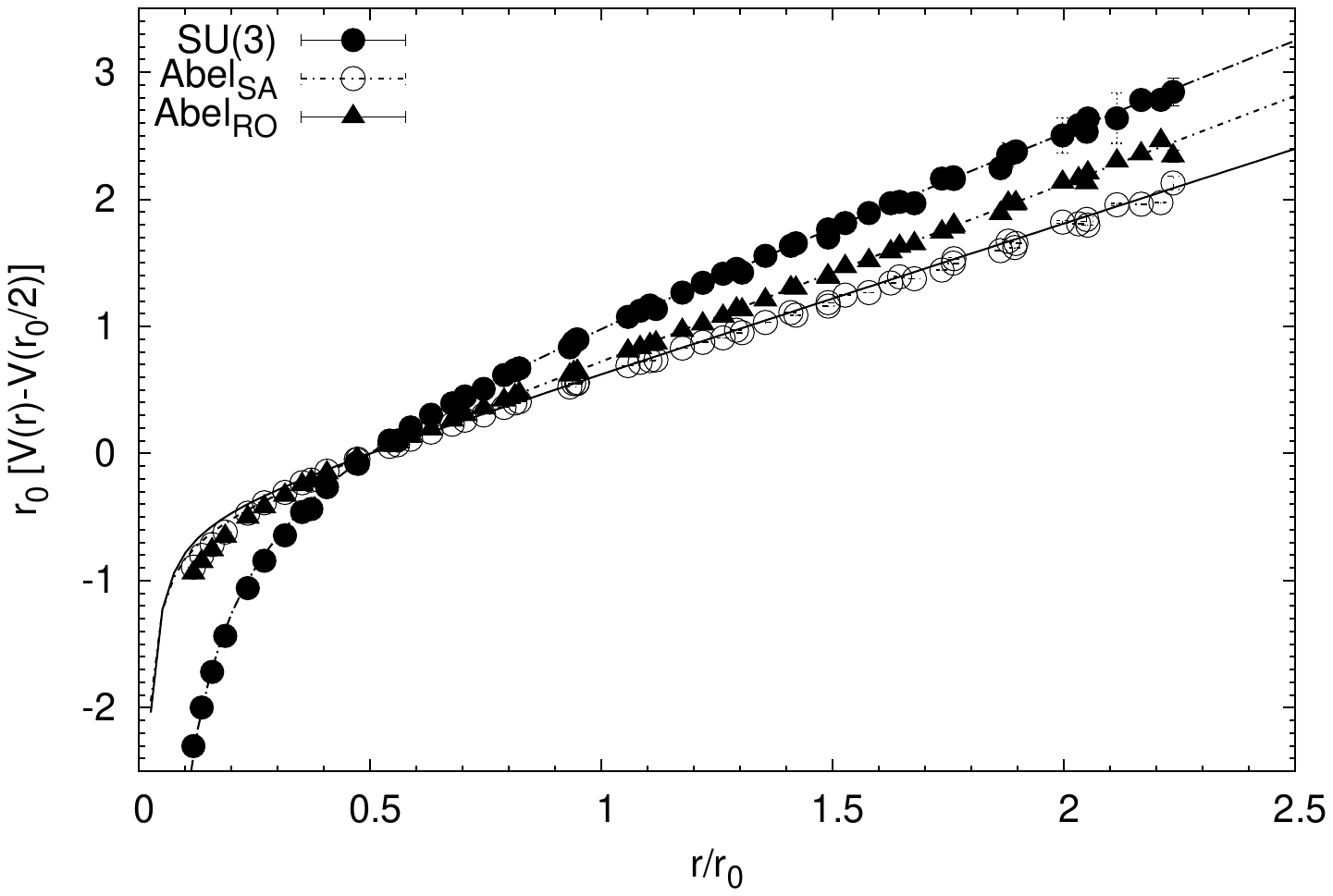} 
\vspace{-10mm}
\caption{Nonabelian static potential (filled circles) is compared with the Abelian static potentials computed for Gribov copies obtained with SA algorithm (empty circles) and with RO algorithm (triangles). The curves show fits of data obtained for minimal lattice spacing (at $\beta=6.3$) to the Cornell potential.} 
\end{center}
\labelf{poten_compar}
\vspace{-5mm}
\end{figure}

\section{Monopole dominance}
Monopoles are defined after Abelian plaquette $ \theta_{\mu \nu}^{a}(x) =  \partial_{\mu} \theta_{\nu}^{a}(x) - \partial_{\nu} \theta_{\mu }^{a}(x) $ is decomposed into regular and singular parts:

\beq
\theta_{\mu \nu}^{a}(x) = \bar \theta_{\mu\nu}^{a}(x) + 2\pi m_{\mu\nu}^{a}(x)
\eeq

Then the monopole currents are defined as:

\beq
j_{\mu}^{a}(s) = \frac{1}{2} \epsilon_{\mu\nu\alpha\beta} \partial_{\nu} m_{\alpha\beta}^{a}(x) = -\frac{1}{4\pi} \epsilon_{\mu\nu\alpha\beta} \partial_{\nu} \bar \theta_{\mu\nu}^{a}(x), ~~~ s_\nu=x_\nu+\half 
\eeq
They  form closed loops due to conservation law $\partial_{\mu} j_{\mu}^{a}(s)=0$ .

Monopole component of the Abelian gauge field is defined as:

\beq
\theta^{mon, a}_{\mu}(x) = -2 \pi \sum_{y} D(x-y)\partial_{\nu}^{'} m_{\nu\mu}^{a}(y)\,,
\eeq
where $\partial_{\nu}^{'}$ is a backward derivative and $D(x)$ is a lattice inverse Laplacian.

Thus the Abelian field $U^{abel}_{\mu}(x)$ can be decomposed into the monopole and photon components in the same way as it is made in compact $U(1)$ gauge theory. The dominance
of this monopole component  $U^{mon}_{\mu}(x)$ in the infrared observables was supported by many observations, see
reviews mentioned above. In particular, it was found that the respective monopole string tension $\sigma_{mon}$ determined from the Wilson loops computed for the gauge field $U^{mon}_{\mu}(x)$ is close to but lower than the physical string tension $\sigma$. The dependence of $\sigma_{mon}$ on the Gribov copies has not been carefully studied so far. Here we close this gap at least partially.

\begin{figure}[t]
\vspace{-5mm}
\begin{center}
\includegraphics[width=0.65\textwidth]{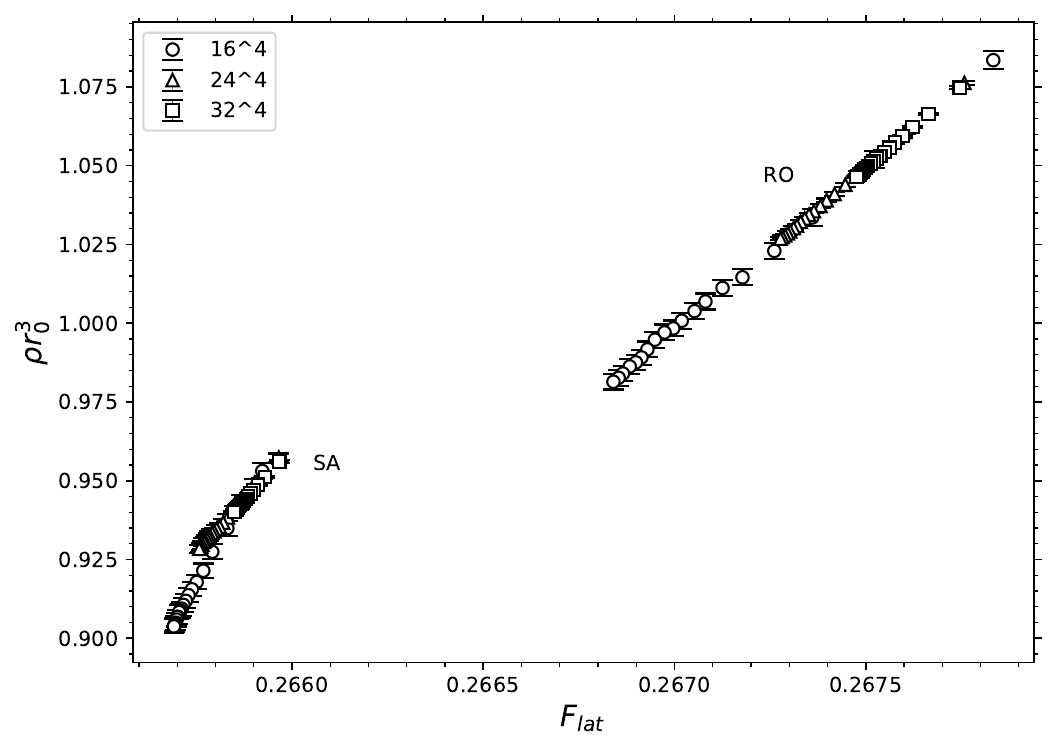}
\caption{Monopole density computed for lattices with $L/a=16, 24, 32$ and lattice spacing $a=0.093 $~fm for Gribov copies obtained with SA algorithm (data ponts in the left part of the figure) and with RO algorithm (data points in the right part).} 
\end{center}
\labelf{mondens}
\end{figure}
In Fig.~\ref{mondens} we show dependence of the monopole density on the value of the functional (\ref{functional2}) obtained using RO and SA algorithms. Each point is obtained by taking average over best (i.e. with minimal value of $F_{lat}$) out of $n$ copies, $n=1,...,20$ both for density and for the functional. The data are presented for one lattice spacing ($a=0.093$~fm) and three lattice sizes: $L/a = 16, 24, 32$. One can see that results, obtained with RO and SA algorithms differ very much. More detailed studies are necessary to undestand this difference. At the same time for given algorithm results obtained for different lattices fall on a universal curve. Furthermore, on smaller lattices it is easier to reach lower value of  $F_{lat}$. The data clearly indicate that the density depends on the value of the functional $F_{lat}$ rather than  on the lattice size. This conclusion is true for other gauge dependent observables. 


 In Fig.~\ref{poten_compar_monop} we compare the static potentials $V_{mon}(r)$ computed for the monopole component  $U^{mon}_{\mu}(x)$ obtained either for one gauge copy with RO algorithm or for $n=20$ gauge copies with SA algorithm, i.e. for same gauge copies as 
 were used in Fig.~\ref{poten_compar} to compute the static potentials $V_{abel}(r)$. Results are shown for all four values of the lattice spacing in the case of RO algorithm and for one (minimal) lattice spacing for SA algorithm. 
The potential demonstrates again a weak dependence on the UV cutoff.
The potentials were fitted to the Cornell potential and the ratio $\sigma_{mon}/ \sigma$ was obtained equal to 0.86(4) for SA algorithm and to 1.01(4) for RO algorithm.

\begin{figure}[t]
\vspace*{-1.5cm}
\begin{center}
\includegraphics[width=0.8\textwidth]{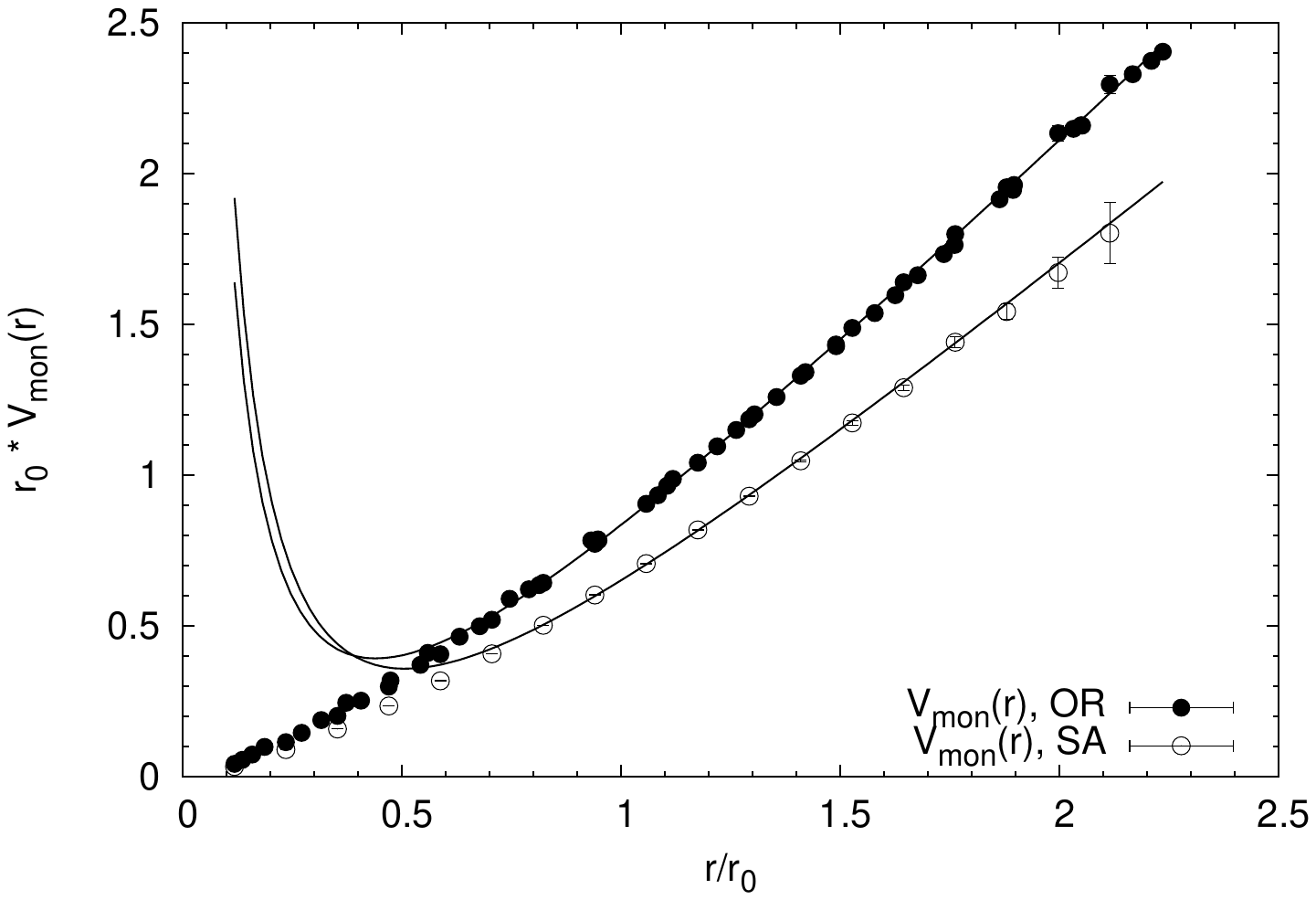}
\vspace{-10mm}
\caption{Monopole static potential $V_{mon}(r)$ computed for Gribov copies obtained with SA algorithm (empty circles) and with RO algorithm (filled circles). The curves show fits of data obtained for minimal lattice spacing (at $\beta=6.3$) to the Cornell potential.} 
\end{center}
\labelf{poten_compar_monop}
\vspace{-5mm}
\end{figure}

\section{Conclusions}
We presented new results on the Gribov copies effects in MA gauge of $SU(3)$  gluodynamics.
 To fix the MA gauge we used two algorithms which produce significantly different values for the gauge fixing functional (\ref{functional2}) as can be seen from Fig.~\ref{mondens}. After fixing the gauge we performed an Abelian projection and also determined the monopole component of the Abelian gauge field. The static potential and respective string tension were computed for  the original gauge field $U_\mu(x)$, for the Abelian gauge field $U_\mu^{Abel}(x)$ and for the monopole component $U_\mu^{mon}(x)$. 
 
 As demonstrated in Fig.~\ref{poten_compar} (see also \cite{Kudrov2024}) on the Gribov copies obtained using SA algorithm the slope of $V^{ab}(r)$ is significantly lower than the slope of $V(r)$ with ratio $\sigma_{ab}/\sigma =0.83(2)$. At the same time the Gribov copies obtained using less effective RO algorithm produce  value compatible with 1: $\sigma_{ab}/\sigma =0.96(3)$. It is worth to note that these results only slightly depend on the lattice spacing. The main message following from this observation is that it is possible to choose Gribov copies which
 demonstrate 'perfect' Abelian dominance \cite{Sakumichi:2014xpa}. 

 Then we turned to the monopole dominance. First,  we presented in Fig.~\ref{mondens}
 dependence of the magnetic currents density vs. the value of the functional $F_{lat}$  for 20 randomly chosen Gribov copies for each of these two algorithms.
 Results were obtained for fixed lattice spacing $a=0.093$~fm and three lattice sizes. The results presented in this figure underline once more the strong difference between Gribov copies obtained with use of SA and RO algorithms. Additionally, one can see that for given algorithm the density depends on value of $F_{lat}$ and is only weakly dependent on
 lattice size. We observed similar weak dependence on the lattice size for other
 gauge dependent observables when they are compared at fixed value of $F_{lat}$.
 This observation confirms our conclusion made in \cite{Kudrov2024} about weak volume dependence for $\sigma_{ab}$. 

 Our results for the static potentials $V_{mon}(r)$ presented in Fig.~\ref{poten_compar_monop} show that dependence on Gribov copies for
 respective string tension $\sigma_{mon}$ is very similar to that of $\sigma_{ab}$: for SA algorithm we obtained $\sigma_{mon}/\sigma=0.86(4)$
 while for OR algoritm  $\sigma_{mon}/\sigma=1.01(4)$ was found, i.e. strong dependence on the Gribov copy choice was demonstrated for this observable. It is important that we found  $\sigma_{mon}/\sigma \approx 1$ on the same Gribov copies on which we found $\sigma_{ab}/\sigma \approx 1$.


\label{sec:acknowledgement}
\section*{Acknowledgement}
Computer simulations were performed on the FEFU GPU cluster Vostok-1,
the Central Linux Cluster of the NRC “Kurchatov Institute”-IHEP, and the Linux Cluster of  KCTEP NRC “Kurchatov Institute”. Resources of the federal collective usage center Complex for Simulation and Data Processing for Mega-Science Facilities at NRC ”Kurchatov Institute” were also used.

\label{sec:funding}
\section*{Funding}
This research was funded by the Russian Science Foundation (Grant 23-12-00072). 
The work of V.B. (generation of lattice gauge field configurations on the FEFU GPU cluster Vostok-1)  is partially supported by the Ministry of Science and High Education of Russia (Project No.~FZNS-2024-0002).


\bibliographystyle{pepan}
\bibliography{bibliography.bib}
\end{document}